\documentclass[conference]{IEEEtran}
\IEEEoverridecommandlockouts
\usepackage{cite}
\usepackage{amsmath,amssymb,amsfonts}
\usepackage{algorithmic}
\usepackage{graphicx}
\usepackage{textcomp}
\usepackage{xcolor}
\usepackage{amsmath,amsfonts}
\usepackage{algorithmic}
\usepackage{algorithm}
\usepackage{array}
\usepackage[caption=false,font=normalsize,labelfont=sf,textfont=sf]{subfig}
\usepackage{textcomp}
\usepackage{stfloats}
\usepackage{url}
\usepackage{verbatim}
\usepackage{graphicx}
\usepackage{cite}
\def\BibTeX{{\rm B\kern-.05em{\sc i\kern-.025em b}\kern-.08em
T\kern-.1667em\lower.7ex\hbox{E}\kern-.125emX}}
\begin{document}

\title{Energy Efficiency Optimization of Intelligent Reflective Surface-assisted Terahertz-RSMA System}
\author{\IEEEauthorblockN{Xiaoyu Chen} %1\textsuperscript{st} Xiaoyu Chen}
\IEEEauthorblockA{\textit{School of Electrical and Information Engineering} \\
\textit{The University of Sydney}\\
Sydney, Australia \\
xche7612@uni.sydney.edu.au}
\and
\IEEEauthorblockN{Feng Yan} %2\textsuperscript{nd} Feng Yan}
\IEEEauthorblockA{\textit{National Mobile Communications Research Laboratory} \\
\textit{Southeast University}\\
Nanjing, 210096, China \\
feng.yan@seu.edu.cn}
\and
\IEEEauthorblockN{Menghan Hu} %3\textsuperscript{rd} Menghan Hu}
\IEEEauthorblockA{\textit{School of Electrical and Information Engineering} \\
\textit{The University of Sydney}\\
Sydney, Australia \\
mehu4638@uni.sydney.edu.au}
\and
\IEEEauthorblockN{Zihuai Lin} %4\textsuperscript{th} Zihuai Lin}
\IEEEauthorblockA{\textit{School of Electrical and Information Engineering} \\
\textit{The University of Sydney}\\
Sydney, Australia \\
zihuai.lin@sydney.edu.au}
%\and
}

\maketitle
% \author{\IEEEauthorblockN{\textsuperscript{} Xiaoyu Chen\textsuperscript{1}, Feng Yan\textsuperscript{2}, Menghan Hu\textsuperscript{1}, Zihuai Lin\textsuperscript{1}}
% \IEEEauthorblockA{\textsuperscript{1}
% \textit{School of Electrical and Information Engineering}, \textit{The University of Sydney}, Sydney, Australia} 
% \IEEEauthorblockA{\textsuperscript{2}
% \textit{National Mobile Communications Research Laboratory}, \textit{Southeast University}, Nanjing, 210096, China }

% Email:\textsuperscript{1}\lbrace {xche7612}, {mehu4638} \rbrace @uni.sydney.edu.au, \textsuperscript{2}\lbrace{feng.yan} \rbrace @seu.edu.cn, \textsuperscript{1}\lbrace {zihuai.lin} \rbrace @sydney.edu.au
% }

% \maketitle

\begin{abstract}
%This paper discusses the energy efficiency of the downlink system of the intelligent reflective Surface (IRS) assisted multiuser Rate Split Multiple Access (RSMA) under terahertz propagation.
%This paper examines the energy efficiency of the intelligent reflective Surface (IRS) assisted multi-user Rate Split Multiple Access (RSMA) downlink system under terahertz propagation. RSMA is a general and robust multiple access framework in multi-antenna systems, which can improve spectral efficiency and enhance the Quality of Service (QoS). In this paper, we use the Salp Swarm Algorithm (SSA) to optimize the objective function of energy efficiency.% The simulation results show that SSA can improve energy efficiency better than Successive Convex Approximation (SCA) technique, and the time consumed is far less than SCA.
%The simulation results reveal that SSA improves energy efficiency more than the Successive Convex Approximation (SCA) technique, and it takes significantly less time.
This paper examines the energy efficiency optimization problem of intelligent reflective surface (IRS)-assisted multi-user rate division multiple access (RSMA) downlink systems under terahertz propagation. The objective function for energy efficiency is optimized using the salp swarm algorithm (SSA) and compared with the successive convex approximation (SCA) technique. SCA technique requires multiple iterations to solve non-convex resource allocation problems, whereas SSA can consume less time to improve energy efficiency effectively. The simulation results show that SSA is better than SCA in improving system energy efficiency, and the time required is significantly reduced, thus optimizing the system's overall performance.

\end{abstract}


\section{Introduction}
%As the wireless world moves towards the sixth generation (6G), there is an urgent need to develop new spectrum resources for future 6G communication systems. Due to the abundant bandwidth in the terahertz (THz) frequency band (between 0.1THz and 10THz), terahertz communication is expected to meet the rate requirements of next-generation wireless communication networks.
%With the advent of the sixth generation, the rate requirements of communication networks are gradually increasing, and the rich bandwidth of the terahertz band can meet the rate requirements of 6G communication networks.
The rate requirements of communication networks are continuously increasing with the advent of the sixth generation (6G), and the rich bandwidth of the terahertz band can meet the rate requirements of 6G communication networks. However, terahertz signals can be affected by obstacles resulting in high reflective attenuation, severely impairing signal coverage and causing degraded communication performance. %Therefore, it is challenging to realize beam steering and beam tracking for terahertz communication. 
This paper incorporates an intelligent reflective surface (IRS) to create a controllable propagation environment \cite{du2022ITS_RIS,chu2022IoT_RIS,Hu2021CodedRIS}. %The central processor typically controls the IRS, and each reflective element can manipulate and change the amplitude and phase shift of the reflection of the impinging terahertz wave.
IRS used in terahertz communication systems can mitigate the fading of wireless channel in high frequency band \cite{xu2021THzUAV,xu2022THzRIS,Liu2022THzRIS}. Through the deployment of IRS, the blocking problem in the light-of-sight (LoS) link can be well avoided \cite{1}. Applying IRS as a phase-shifting structure to the antenna array signal adjusts the beam direction by changing the electromagnetic properties of the electromagnetic wave, which helps to sharpen the beam shape and thus affects the surrounding propagation environment. Therefore, optimizing the beamforming to improve the system's energy efficiency at a reasonable cost is essential.

To meet the demand for high energy efficiency and quality of service (QoS) heterogeneity in wireless networks, rate split multiple access (RSMA) is developed in \cite{2}. RSMA is considered as an indispensable technology to support large-scale connections of 6G wireless communication networks. Space-Division Multiple Access (SDMA) uses linear precoding to separate users in the spatial domain and treats any residual multi-user interference entirely as noise. Non-Orthogonal Multiple Access (NOMA) uses linear precoding superposition coding with sequential interference cancellation (SIC) to superimpose users in the power domain. It relies on user grouping and forces some users to decode and eliminate interference from other users. RSMA is a multiple access framework that differs from SDMA and NOMA in that RSMA is more versatile \cite{2} as it partially decodes the interference and treats the rest as noise. Recently, in\cite{3}, it is shown that RSMA can be the preferred choice in situations when neither SDMA nor NOMA is applicable, and RSMA can achieve superior performance compared to NOMA and SDMA under different precoding methods. In \cite{4}, it demonstrates that RSMA is superior to SDMA for different algorithms. Furthermore, RSMA outperforms SDMA and NOMA over a wide range of network loads (underload and overload conditions) and user deployment rates.

%In \cite{4}\cite{5}, the application of RSMA in different scenarios is mentioned. For example, in fast user movement scenarios, RSMA can be found to optimise the user rate with shorter codes, solving problems in low latency and enhanced mobile broadband (eMBB).
The applications of RSMA in various scenarios are discussed in \cite{4}\cite{5}. In fast user movement scenarios, for example, RSMA can be used to optimize the user rate with shorter codes, hence eliminating difficulties with reduced latency and enhanced mobile broadband (eMBB). In \cite{6}, the application of RSMA on multiple-input single-output (MISO) is extended to multiple-input multiple-output (MIMO) by proposing to optimise the RS pre-coder using the weighted Minimum Mean Square Error (WMMSE) and Sample Average Approximation (SAA) such that WSR (weight sum-rate) can be constrained, and the results show that RSMA outperforms NOMA in the MIMO case. Combining RSMA and IRS has also become one of the recent research directions. In \cite{7}, it is mentioned that combining IRS and RSMA can make the baseline scheme more optimised, and a combination of successive convex approximation (SCA) and alternating optimization (AO) algorithms is proposed to optimise the performance. After the optimisation, it can be found that RSMA can effectively allocate resources fairly to achieve the increase in users' rate. In \cite{8}, a comparison of RSMA with SDMA and NOMA for energy efficiency 
(EE) maximization in multiple input single output broadcast channels (MISO BC) is presented. However, the SCA approach is relatively iteratively complex, computationally intensive and generally effective. Therefore, in this paper, the salp swarm algorithm (SSA) is used for optimization.

%The contributions of this paper are:
The main contributions of this paper as follows:

Firstly, a new IRS-assisted terahertz RSMA framework is proposed.

Secondly, this paper uses the Salp Swarm Algorithm (SSA) to optimise energy efficiency problems under power and quality of service constraints. 

Finally, simulation results show that SSA is superior to Successive Convex Approximation (SCA) in energy efficiency optimization, and the time complexity is significantly lower than SCA.
%Section \uppercase\expandafter{\romannumeral2} describes the structure of the terahertz communication system, RSMA and system model in detail. Section \uppercase\expandafter{\romannumeral3} describes optimisation of energy efficiency problems and the SSA algorithm. %The proposed energy efficiency problem of RSMA is discussed in Section \uppercase\expandafter{\romannumeral4}, followed by the proposed algorithm based on SSA in Section \uppercase\expandafter{\romannumeral4}.
%Section \uppercase\expandafter{\romannumeral4} compares SSA and SCA for maximum energy efficiency and time complexity, and Section \uppercase\expandafter{\romannumeral5} summarizes the paper.

The outline of the paper is given below. Section II elaborates on the terahertz communication system's structure, RSMA, and system model. Optimization of energy efficiency issues and the SSA technique are described in Section III. The maximum energy efficiency and time complexity of SSA and SCA are compared in Section IV, the paper is concluded in Section V.

\section{IRS assisted Terahertz-RSMA system}
\subsection{Terahertz Communication Channel Model}
%Terahertz communication has a high transmission rate, large capacity, strong directivity, high security, and very poor penetration ability. It need to mentioned that terahertz propagates in the air, it is easily absorbed by moisture, and the signal is severely attenuated, which leads to high-frequency selective path loss of the line-of-sight (LOS) link. For the non-line-of-sight (NLOS) propagation, it depends on the shape, material, and roughness of the reflecting surface. In \cite{9}, the LOS propagation model in the whole terahertz band is established by analyzing the influence of molecular absorption on path loss. As described in \cite{10}, the NLOS propagation model reflects the loss and includes the influence of molecular absorption and propagation loss on the indirect path. 

This paper proposes a terahertz communication channel model based on Kirchhoff scattering theory, including the effect of line-of-sight propagation. The line-of-sight propagation model proposed in \cite{9} is improved by using the updated HITRAN 2012 database and the actual value of water vapor content in the air. Path loss refers to the channel link attenuation caused by the propagation environment between the transmitter and receiver. In terahertz path loss $H^{LOS}(f,r)$ is composed of propagation loss $H_{spread}(f,r)$ and molecular absorption loss $H_{abs}(f,r)$, which are shown as follows:
\begin{equation}
H^{LOS}(f,r)=H_{spread}(f,r)\cdot H_{abs}(f,r).
\end{equation}
The propagation loss is shown as follows:
\begin{equation}
H_{spread}(f,r)=(\frac{4\pi fr}{c})^{2},
\end{equation}
where $c=3\times 10^{8}$, representing the speed of light; $f$ is the operating frequency; $r$ is the distance between the transmitter and receiver.

Molecular absorption loss $H_{abs}(f,r)$ is expressed as follows:
\begin{equation}
H_{abs}(f,r)=e^{-\frac{1}{2}\mathcal{K}_{a}(f)r},
\end{equation}
where $\mathcal{K}_{a}(f)$ is the medium absorption coefficient of the relative area under the unit volume.

To calculate $\mathcal{K}_{a}(f)$, this paper adopts a simplified molecular absorption coefficient model for 275GHz frequency band \cite{10}, which is shown as follows:
\begin{equation}
\mathcal{K}_{a}(f)=x(f,v)+y(f,v)+z(f,v),
\end{equation}
\begin{equation}
x(f,v)=\frac{(q_{1}v(q_{2}v+q_{3}))}{(q_{4}v+q_{5})^{2}+(\frac{f}{100c}-p_{1})^{2}},
\end{equation}
\begin{equation}
y(f,v)=\frac{(q_{6}v(q_{7}v+q_{8}))}{(q_{9}v+q_{10})^{2}+(\frac{f}{100c}-p_{2})^{2}},
\end{equation}
\begin{equation}
z(f,v)=c_{1}f^{3}+c_{2}f^{2}+c_{3}f+c_4,
\end{equation}
where $q_{1}=0.2205$, $q_{2}=0.1303$, $q_{3}=0.0294$, $q_{4}=0.4093$, $q_{5}=0.0925$, $q_{6}=2.014$, $q_{7}=0.1702$, $q_{8}=0.0303$, $q_{9}=0.537$, $q_{10}=0.0956$, $c_{1}=5.54\times 10^{-37}Hz^{-3}$, $c_{2}=-3.94\times 10^{-25}Hz^{-2}$, $c_{3}=9.06\times 10^{-14}Hz$, $c_{4}=-6.36\times 10^{-3}Hz^{3}$, $p_{1}=10.835\times 10^{-2}cm^{-1}$, $p_{2}=12.664\times 10^{-2}cm^{-1}$. $v=\frac{\rho p_{w}(T,P)}{100p}$ is the mixing ratio of water vapor in unit volume, where $\rho$ and $v$ are relative humidity and atmospheric pressure, respectively, where $\rho =0.5$ and $p=101325$. $p_{w}(T,p)$ is the partial pressure of saturated water vapor at temperature $T$, expressed as follows:
\begin{equation}
p_{w}(T,p)=g_{1}(g_{2}+g_{3}\eta_{h})e^{\frac{g_{4}(T-g_{5})}{T-g_{6}}},
\end{equation}
where $g_{1}=6.1121$, $g_{2}=1.0007$, $g_{3}=3.46\times 10^{-6}$, $g_{4}=17.502$, $g_{5}=273.15^{\circ}K$, $g_{6}=32.18^{\circ}K$, $T=296$, $\eta_{h}$ is the pressure, which is 1013.15 hPa\cite{10}.
%$\eta_{h}=1013.15$ pressure is equal to 1013.15 hPa. 

The equivalent channel transfer function is given by: 
\begin{equation}
H(f,r)=H^{LOS}(f,r)e^{-j2\pi f\tau_{LOS}},
\end{equation}
where $\tau_{LOS} = r/c$ is the propagation delay of the LOS path. 

\subsection{Rate Split Multiple Access}
%Power-domain non-orthogonal multiple access (NOMA) based on transmitter-side superposition coding (SC) and receiver-side sequential interference cancellation (SIC) is currently used in future mobile networks\cite{9}. While NOMA can serve users in overload situations, it is only suitable for deployments where user channels are well aligned, and their strengths vary widely\cite{12}. %Multi-antenna NOMA requires one user to decode other users' messages fully. But this design constraint results in loss of multiplexing gain, rate, spectral efficiency, inability to serve many users, and inefficiencies in using SIC receivers in multi-antenna setups.
To overcome the two extreme interference management strategies of SDMA and NOMA, RSMA is a multiple access scheme for multi-antenna multi-user communication based on rate splitting (RS) and linear precoding. RSMA splits the messages of users into public and private parts and encodes the public part into one or more public streams while encoding the private part into a separate private stream. These streams are precoded using available perfect or imperfect channel state information, superimposed, and transmitted via MIMO or MISO BC. All receivers first decode the common stream, perform SIC, and then decode their respective private streams. Each receiver reconstructs its original message from the message part in the public stream and its expected private stream. The main advantage of RSMA is that it flexibly manages interference, allowing it to be partially decoded and partially treated as noise. RSMA's ability to decode partial interference and partial interference as noise softly bridges the extreme cases of fully decoding interference \cite{11} and treating interference as noise and provides room for increased rate and QoS and reduced complexity.
\begin{figure}[h]
\centering
\includegraphics[width=3in]{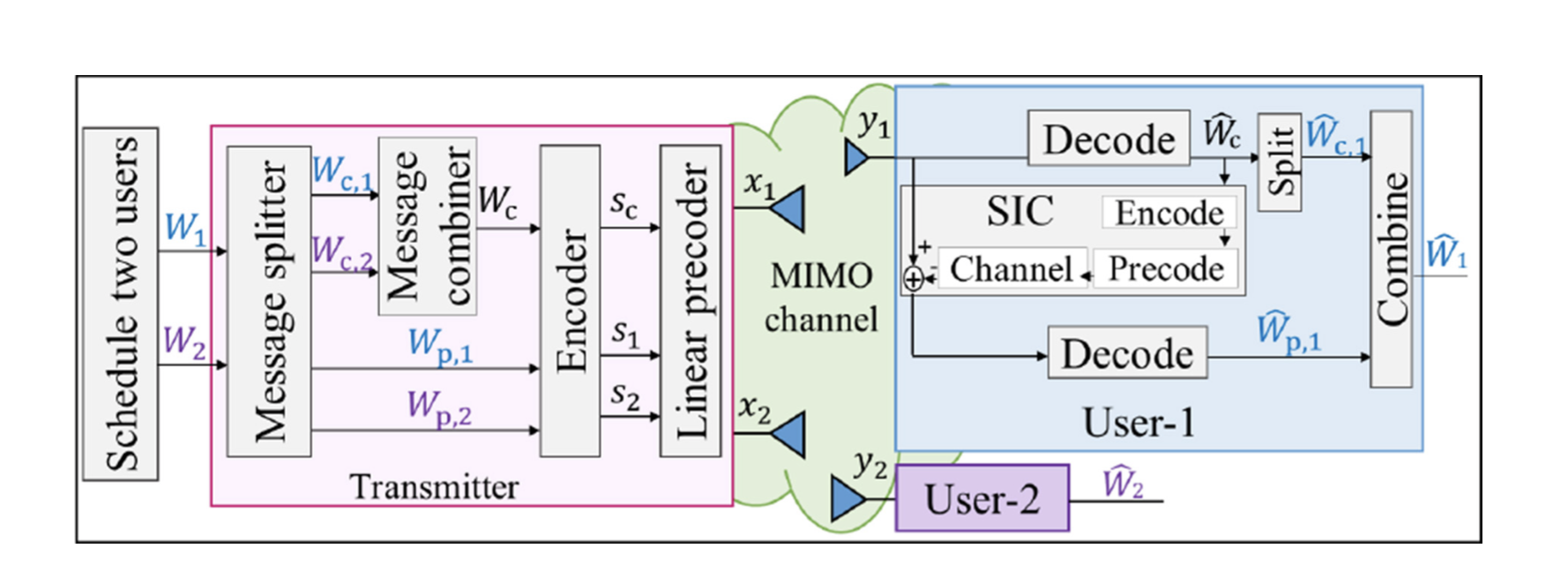}
\caption{Two-user system architecture with rate-splitting}
\label{Two-user system architecture with rate-splitting}
\end{figure}
When two users are in the system, the generalized RSMA proposed in \cite{2} reduces to the 1-layer RS. % studied in [11]. 
The main difference between the two types of RSMA is the generation of the transmit signal $x$. RSMA's message $W_k, \forall k \in \lbrace 1, 2\rbrace$ is split into a common part ${W}_{c,k}$ and a private part ${W}_{p,k}$. The common parts of the two users, $W_{c,1}, W_{c,2}$ are jointly encoded into a common stream $s_c$ using a codebook shared by the two users. The common stream $s_c$ works for both users. The private part is encoded as $s_1$ and $s_2$ of user-1 and user-2, respectively. Stream vector $\mathbf{s} = [s_c,s_1,s_2]^T$ linearly precoded using beamformer $\mathbf{P} = [\mathbf{P}_c, \mathbf{P}_1, \mathbf{P}_2]$. 

Therefore, the BS sends a superimposed signal stream containing both a common message stream and $K$ private message streams for $K$ users.
\begin{equation}
\mathbf{x}=\underbrace{\mathbf{P}_{c}s_c}_{Common\quad message}+\underbrace{\sum_{k=1}^{K}\mathbf{P}_{k} s_{k}}_{Private\quad message}, \forall k \in \lbrace 1,2,...,K \rbrace.
\end{equation}
%The transmit power $\mathbb{E}\lbrace {\mid s_c\mid}^2 \rbrace =\mathbb{E}\lbrace {\mid s_k\mid}^2 \rbrace =1, \forall k$ is also bounded by $P_t$.
The total transmit power of the base station is affected by the power constraint $P_t$, that is, $\mathbb{E}\lbrace \lVert x \rVert ^2 \rbrace \leq P_t$.

\subsection{System Model}
\begin{figure}
\centering
\includegraphics[width=3in]{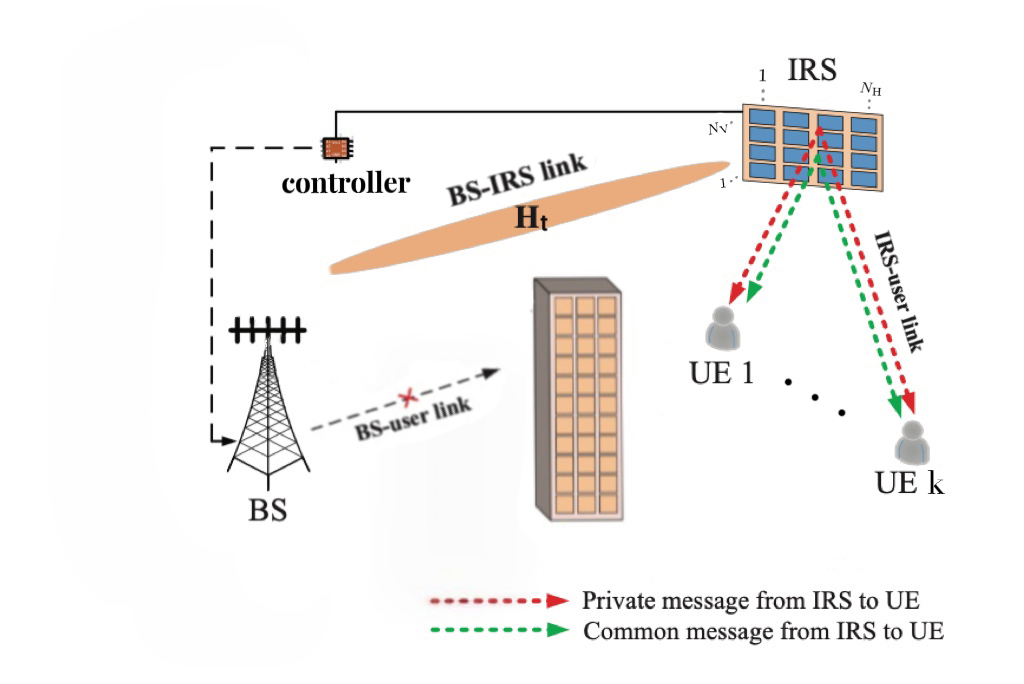}
\caption{An IRS-aided downlink MISO Terahertz-RSMA system.}
\label{An IRS-aided downlink MU-MISO RSMA system with one BS, one IRS, and K users.}
\end{figure}

As shown in Figure 2, an IRS-assisted multi-user MISO with RSMA downlink system is considered in this paper. BS is equipped with $N_t$ transmit antennas, serving two single-antenna users. The direct path is the channel response of the base station to the user terahertz communication channel. When considering the direct path has a strong attenuation in the high-frequency band, causing the receiver to be unable to receive effectively, the signal of the direct path is not analyzed. IRS is assumed to be a uniform planar array (UPA) with $N_{IRS} = N_H \times N_V$ passive reflective elements, which are deployed on a two-dimensional rectangular grid with $N_H$ elements per row and $N_V$ elements per column, and the height of the IRS from the ground is $H$\cite{12}.

The coordinates of UE $k$ is denoted by $w_{k} = [x_k, y_k, z_k]^T$, and the coordinates of the first reflective element are denoted by $\mathbf{l}_1 = [X, Y, H]^T$, where $T$ represents the matrix transpose. So the coordinates of the $n$-th reflective element is $\mathbf{l}_n=[X+(n_{x}-1)\Delta,Y +(n_{y}-1)\Delta,H]^T$, where $n=(n_{y}-1)N_{V} +n_x,n_x =1,. ..,N_H$,and $n_{y}=1,...,N_V$,  and the separation between adjacent IRS elements is denoted by $\Delta$.

The phase $\alpha_{N_{IRS}}(f, \mathbf{l})=\frac{2\pi f\mathbf{r}_{t}(\mathbf{l})^{T}}{c \mid \mathbf{r}_{t}(\mathbf{l})\mid}(\mathbf{l}_{n}- \mathbf{l}_1)$ represents the phase difference of the RF signal of frequency $f$ entering the $n$-th reflective element relative to the first element, where 
$\mid \mathbf{r}\mid$ represents the Euclidean norm of vector $\mathbf{r}$; $\mathbf{r}_{t}= [X-x_t, Y-y_t, H-z_t]^T$ denote the distance of BS-IRS; $c$ is the speed of light.

The coordinates of BS are represented by $w_{t} = [x_t, y_t, z_t]^T$. The transmit steering vector of the BS-IRS link is defined as 
\begin{equation}
\mathbf{e}_{t}(f,\mathbf{l}) = [1, e^{-j\alpha_{1}(f,\mathbf{l})},..., e^{-j\alpha_{N_{IRS}}(f,\mathbf{l})}]. 
\end{equation}

The phase difference of the RF signal of frequency $f$ received from the nth reflective element at UE $k$ relative to the first reflective element is $\phi_{N_{IRS}}(f, \mathbf{l})=\frac{2\pi f\mathbf{r}_{k}(\mathbf{l})^{T}}{c \mid \mathbf{r}_{k}(\mathbf{l})\mid}(\mathbf{l}_{n}- \mathbf{l}_1)$, where $\mathbf{r}_{k}= [x_k-X, y_k-Y, z_k-H]^T$ denotes the distance of the IRS-user $k$ links.
The link from IRS to UE $k$ defines the receive steering vector as
\begin{equation}
\mathbf{e}_{k}(f,\mathbf{l}) = [1, e^{-j\phi_{1}(f,\mathbf{l})},..., e^{-j\phi_{N_{IRS}}(f,\mathbf{l})}].
\end{equation}

According to\cite{13}, the channel attenuation of the BS-IRS-user $k$ link is:
\begin{equation}
h_k(f,\mathbf{l})=\frac{\sqrt{G_t}\sqrt{G_r}c}{8\sqrt{\pi^{3}}f\mid \mathbf{r}_{t}(\mathbf{l})\mid \mid \mathbf{r}_{k}(\mathbf{l})\mid}e^{-\frac{1}{2}\mathcal{K}_{a}(f) \mid \mathbf{r}(\mathbf{l})\mid}e^{-j2\pi f\tau_{LOS}},  
\end{equation}
where $G_t$ is transmission antenna gain and $G_r$ is receiving antenna gain; $\mid \mathbf{r}(\mathbf{l})\mid=\mid \mathbf{r}_{t}(\mathbf{l})\mid+\mid \mathbf{r}_{k}(\mathbf{l})\mid$; $\tau_{LOS} = \mid \mathbf{r}(\mathbf{l})\mid/c$ is the propagation delay of the LOS path.

Therefore, the effective combined end-to-end channel from the BS to users with the existence of the IRS can be expressed as:
\begin{equation}
\mathbf{h}_{k}^{H}=\underbrace {h_k(f,\mathbf{l}) \mathbf{e}_{k}(f,\mathbf{l})^{H}\mathbf{\Phi}\mathbf{e}_{t}(f,\mathbf{l})}_{BS-IRS-user}.
\end{equation}
%where $\mathbf{h}_{r}\in \mathbb{C}^{N_T\times 1}$, $\mathbf{H}_{t} \in \mathbb{C}^{N_{IRS}\times N_T}$ denote the IRS-user link, and the BS-IRS link, respectively. The IRS consists of $N_{IRS}$ passive reflecting elements. 
The configuration of the IRS is determined by the diagonal phase-shift matrix $\mathbf{\Phi}=diag(\beta_{1}e^{-j\theta_1},...,\beta_{N_{IRS}}e^{-j\theta_{N_{IRS}}})$, where $\beta$ is amplitude modulation, $\theta$ denote the phase shift of the $n$-th reflecting element of the IRS, which can be carefully adjusted by an IRS controller.

This paper considers that the phase shift of each element of the IRS can only take a finite number of discrete values obtained through the uniform quantization interval $[0,2\pi)$. Therefore, the following equation gives the set of discrete phase shift values of each element:
\begin{equation}
\mathcal{F}=\lbrace 0,\frac{2\pi}{2^{b}},...,\frac{2\pi}{2^{b}}(2^{b}-1)\rbrace,
\end{equation}
where constant $b$ denotes the number of bits used to indicate the maximum number of phase shift levels. %The initial phase probability matrix $\mathbf{p}^{0}=\frac{\mathbf{1}_{2^{b}\times N_{IRS}}}{2^{b}}$ of IRS, $\mathbf{1}_{2^{b}}$ is $2^{b}\times N_{IRS}$-dimensional all 1-matrix.

The base station sends a superimposed signal stream containing public message streams and $K$ private message streams for $K$ users. The received signal of user-$k$ is
\begin{equation}
y_k=\mathbf{h}^H_{k}(\mathbf{P}_{c}s_c+\sum_{k=1}^{K}\mathbf{P}_{k} s_{k})+n_k, \forall k \in \lbrace 1,2,...,K \rbrace,
\end{equation}
where $n_k \sim \mathcal{CN}(0,\sigma_{k}^2)$ is additive white Gaussian noise (AWGN) with zero mean and variance $\sigma_{k}^2$ at user $k$.

\section{Optimization Problem Formulation}
\subsection{Problem Formulation}
This paper adopts the linear power model specified in \cite{10}. The common stream $s_c$ is first decoded at both users by treating the interference from the private streams $s_1$ and $s_2$ as noise. Since $s_c$ contains part of the intended message and part of the message that interferes with the user, it is able to decode the partial interference and treat the interfered part as noise. The SINR for decoding the common stream  $s_c$ at user-$k, \forall k \in \lbrace 1, 2\rbrace$ is 

\begin{equation}
\gamma_{ck}(\mathbf{P})=\frac{\mid \mathbf{h}_{k}^{H}\mathbf{P}_c\mid^2}{\mid \mathbf{h}_{k}^{H}\mathbf{P}_1\mid^2+\mid \mathbf{h}_{k}^{H}\mathbf{P}_2\mid^2+\sigma_{k}^2}.
\end{equation}

%The achievable rate of decoding $s_c$ at user-$k$ 
%The decoding rate at user-$k$ is $C_{k}(\mathbf{P})=Wlog_{2}(1+\gamma_{ck}(\mathbf{P}))$. $N_{0k}=W\sigma_{nk}^2$ is the noise power at user-$k$ over the transmission bandwidth. %o ensure that $s_c$ can be successfully decoded, the common rate must not exceed:
%\begin{equation}
%C(\mathbf{P})=min{\lbrace C_{1}(\mathbf{P}),C_{2}(\mathbf{P})\rbrace}.
%\end{equation}

According to the traditional ordered RS decoding system\cite{14},\cite{1}, the achievable rate (bits/sec/Hz) for the common and private messages of user-$k$ is

\begin{equation}
C_{k}=log_{2}(1+\frac{\mid \mathbf{h}_{k}^{H}\mathbf{P}_c\mid^2}{\sum_{k=1}^{K}\mid \mathbf{h}_{k}^{H}\mathbf{P}_k\mid^2+\sigma_{k}^2}), \forall k \in \lbrace 1,2,...,K \rbrace.
\end{equation}
%\begin{equation}
%C_{1}+C_{2}=C(\mathbf{P}),
%\end{equation}

%For any UE, the common actual transmission data rate shall satisfy the following constraints: 
%\begin{equation}
%C(\mathbf{P})=min{\lbrace C_{1}(\mathbf{P}),C_{2}(\mathbf{P})\rbrace}.
%\end{equation}
%\begin{equation}
%C_{k} \geq \sum_{i=1}^{K}r_{ci},\forall k \in \lbrace 1,2 \rbrace,
%\end{equation}
%where $r_{ci}$ is the common rate that split from the total common rate allocated to UE, which imposed to ensure that all UEs can decode the common message.
The total transmission rate of a common message is the sum of the rates of the common parts divided among all UEs. $s_c$ is decoded and removed from the received signal by SIC, and user-$k$ decodes its desired private stream $s_k$ by treating the interference of user-$i(i \neq k)$ as noise. %Decode the SINR of private stream $s_k$ at user-$k,
The SINR for decoding the private stream  $s_k$ at user-$k, \forall k \in \lbrace 1, 2\rbrace$ is
\begin{equation}
\gamma_{k}(\mathbf{P})=\frac{\mid \mathbf{h}_{k}^{H}\mathbf{P}_c\mid^2}{\mid \mathbf{h}_{k}^{H}\mathbf{P}_i\mid^2+\sigma_{k}^2},
\end{equation}
%The achievable rate of decoding $s_k$ at user-$k$ is
so the decoding rate at user-$k$ is %$R_{k}(\mathbf{P})=Wlog_{2}(1+\gamma_{k}(\mathbf{P}))$. 
\begin{equation}
R_{k}=log_{2}(1+\frac{\mid \mathbf{h}_{k}^{H}\mathbf{P}_k\mid^2}{\sum_{i\neq k}^{K}\mid \mathbf{h}_{k}^{H}\mathbf{P}_i\mid^2+\sigma_{k}^2}).
\end{equation}
The achievable rate of user-$k$ is $R_{total}(\mathbf{P})=C_{k}+R_{k}(\mathbf{P})$.
The model of the RSMA-based EE maximization problem is
\begin{equation}
max_{\mathbf{c},\mathbf{P}}\frac{\sum_{k\in \lbrace 1,2 \rbrace} (C_{k}+R_{k}(\mathbf{P}))}{tr(\mathbf{P}\mathbf{P}^{H})+P_{0}}
\end{equation}
\begin{equation}
s.t.
\begin{cases} 
C_1+C_2 \leq C_{k}(\mathbf{P}), \forall k \in \lbrace 1,2 \rbrace \\
\mathbf{c} \geq 0 \\
%0 < \theta_{n} < 2\pi, 1\leq n\leq N_{IRS} \\
\theta_{n}\in \mathcal{F}, \forall n=1,...,N_{IRS} \\
\Vert \mathcal{F} \Vert_{F}^{2}=1 \\
%C_{k} \geq \sum_{i=1}^{K}r_{ci},\forall k \in \lbrace 1,2 \rbrace \\
%r_{ck}\geq 0, \forall k \in \lbrace 1,2 \rbrace, \\
\Vert \mathbf{P}_{c} \Vert^{2}+\sum_{k=1}^{K}\Vert \mathbf{P}_{k} \Vert^{2} \leq P_t,  \forall k \in \lbrace 1,2,...,K \rbrace
\end{cases},
\end{equation}
%\begin{equation}
%s.t \quad C_1+C_2 \leq C_{k}(\mathbf{P}), \forall k \in %\lbrace 1,2 \rbrace
%\end{equation}
%\begin{equation}
%tr(\mathbf{P}\mathbf{P}^{H})\leq P_t   
%\end{equation}
%\begin{equation}
%\mathbf{c} \geq 0
%\end{equation}
%where user 1's weight is fixed at $u_1 = 1$, and the weight of user 2 is changed to $u_2 = 10^{[-3,-1,-0.95,...,0.95,1,3]}$. 
where $P_0$ is the power consumption; $\mathbf{c}=[C_1,C_2]$ is the common rate vector; $P_{t}$ is the maximum transmit power; $\theta$ represents the phase shift of IRS; $\Vert \cdot \Vert_{F}$ is the Frobenius norm.

\subsection{SSA}
The EE maximization problem is non-convex fractional programming. The SCA algorithm is used in the literatures \cite{14}, \cite{15}, \cite{16}, \cite{17} that maximizes EE, but the iterative process is too complicated, so this paper solves EE maximization problem based on SSA. The algorithm simulates the swarm behavior of the salps chain and is a swarm intelligence optimization algorithm. The salps chain is divided into leaders and followers. In each iteration, the leader moves toward the optimal solution of the objective function and guides the movement of the followers, which are only affected by the previous salps. In the iteration, the leader conducts global exploration, while the followers thoroughly conduct local exploration. This method enables the salps chain to have strong global exploration and local development capabilities, significantly reducing the falling into local optimum.

\subsubsection{Initialization}
Let the search space be a Euclidean space of $D\times N$, where $D$ is the space dimension, and $N$ is the number of populations. The position of salps in space is represented by $X_{n}=[X_{n1},X_{n2},...,X_{nD}]^{T}$, and the position of the optimal result is represented by $F_{n}=[F_ {n1},F_{n2},...,F_{nD}]^{T}$ with $n=1,2,3,...,N$. The upper bound of the search space is $ub=[ub_{1},ub_{2},...,ub_{D}]$, the lower bound is $lb=[lb_{1},lb_{2},...,lb_{D}]$.
\begin{equation}
X_{D\times N}=rand(D,N)\dot (ub(D,N)-lb(D,N))+lb(D,N).
\end{equation}
The leader in the population is represented by $X_{d}^{1}$, and the follower is represented by $X_{d}^{i}$, with $i=2, 3,4,...,N$ and $d=1,2,3, ...D$.
Then sort the salps and set the position of the first salp with the best value to the current position. After selecting the location, $N-1$ salps remain in the group. According to the order, the top half is regarded as leaders, and the remaining salps are regarded as followers.

\subsubsection{Leader Position Update}
During the movement and the search for the optimal result  of the salp chain, the leader's position update is expressed as:
\begin{equation}
X_{d}^1=
\begin{cases} 
F_{d}+c_{1}((ub-lb)c_{2}+lb), &c_{3}\geq 0.5 \\
F_{d}-c_{1}((ub-lb)c_{2}+lb), &c_{3} < 0.5
\end{cases},
\end{equation}
where $X_{d}^{1}$ and $F_{d}$ are the position of the first salp (leader) and the optimal result in the $d$-th dimension, respectively; $ub$ and $lb$ are the corresponding upper and lower bounds, respectively. Among them, $c_{1}$, $c_{2}$, $c_{3}$ are control parameters. $c_{1}$ is the convergence factor in the optimization algorithm, which balances global exploration and local development. It is the most important control parameter in SSA. The expression of $c_{1}$ is:
\begin{equation}
c_{1}=2e^{-(\frac{4l}{L})^2},
\end{equation}
where $l$ is the current iteration number, and $L$ is the maximum iteration number. The convergence factor is a decreasing function. The control parameters $c_{2}$ and $c_{3}$ are random numbers of [0,1], which are used to enhance the randomness of $X_{d}^{1}$ and improve the global search and individual diversity of the chain group.

\subsubsection{Follower Location Update}
During the movement and the search for the optimal result of the salp chain, the followers move forward in a chain-like manner through the mutual influence between the front and rear individuals. %Their displacement conforms to Newton's law of motion, and the follower's motion displacement is:
%\begin{equation}
%X=\frac{1}{2}at^{2}+v_{0}t,
%\end{equation}
%where $t$ is the time, $a$ is the acceleration. The calculation formula of $a$ is $a=(v_{final}-v_{0})/t$, $v_{0}$ is the initial velocity, and $v_{fianl}=(X_{d}^ {i}-X_{d}^{i-1})/t$.
%Considering that in the optimization algorithm, t is iterative, let $t=1$ and $v_{0}=0$ in the iterative process. Then formula (33) can be expressed as:
%\begin{equation}
%X=\frac{X_{d}^{i}-X_{d}^{i-1}}{2},
%\end{equation}
%where $i\geq 2$, $X_{d}^{i}$ and $X_{d}^{i-1}$ are the positions of two salps that are closely connected in the $d$-th dimension, respectively. 
Therefore, the position of the follower is expressed as:
\begin{equation}
X_{d}^{i^{'}}=\frac{X_{d}^{i}-X_{d}^{i-1}}{2},
\end{equation}
where $X_{d}^{i'}$ and $X_{d}^{i}$ are the updated follower's position and the pre-update follower's position in the $d$-th dimension, respectively.

\begin{figure}[!t]
%\begin{algorithm}[!h]
 \label{alg:LSB}
 \begin{algorithm}[H]
    \caption{Salp Swarm Algorithm intelligence-based optimization.}
     \begin{algorithmic}[1]
    \STATE Initialize the position of salps $X_{n} (n=1,2,..., N)$, considering upper bound and lower bound are $ub$ and $lb$ respectively.
    \WHILE{$\Vert \mathbf{P}_{c} \Vert^{2}+\sum_{k=1}^{K}\Vert  \mathbf{P}_{k} \Vert^{2} \leq P_t$}
    \STATE Calculate the initial fitness value of $N$ salps according to the upper and lower limits of each search space dimension. \\
    \STATE $F_n$ is the optimal search agent. \\
    \STATE Update $c_{1}$ according to $c_{1}=2e^{-(\frac{4l}{L})^2}$.
   \FOR{each salp $x_{i}$}
    \IF {($n==1$)}
               \STATE Update the leader's position according to \\
$X_{d}^1=\begin{cases} 
F_{d}+c_{1}((ub-lb)c_{2}+lb), &c_{3}\geq 0.5 \\
F_{d}-c_{1}((ub-lb)c_{2}+lb), &c_{3} < 0.5
\end{cases}$.
           \ELSE
                \STATE Update the follower's position according to \\ $X_{d}^{i^{'}}=\frac{X_{d}^{i}-X_{d}^{i-1}}{2}$.
            \ENDIF
        \ENDFOR
            \STATE Limit each salp to the upper and lower bounds of the variable
   \ENDWHILE
   \RETURN $F_n$
  \end{algorithmic}
 \end{algorithm}
\end{figure}

\section{Results and Analysis}
This paper evaluates the algorithm's performance by comparing SSA and SCA algorithms for RSMA to calculate the maximum energy efficiency of two users under the downlink MISO system. The sum of the maximum energy efficiency is the sum of the maximum energy efficiency that can be achieved by a single energy source. The energy efficiency of two users in RSMA is defined as
\begin{equation}
EE_1=\frac{C_{1}+R_{1}(\mathbf{P})}{tr(\mathbf{P}\mathbf{P}^{H})+P_{0}}. 
\end{equation}
\begin{equation}
EE_2=\frac{C_{2}+R_{2}(\mathbf{P})}{tr(\mathbf{P}\mathbf{P}^{H})+P_{0}}. 
\end{equation}

This experiment sets the frequency of the terahertz channel to $f=0.275 THz$. %and set transmission gain $G_t=1 \times 10^5$ and receive gain $G_r =1 \times 10^6$. 
Molecular absorption coefficient is euqal to $2.8\times 10^{-4}$. Assume that the BS has four transmit antennas $N_t = 4$. Noise variance is $\sigma_{nk}=1$ and bandwidth is $W=1Hz$. The transmit power is constrained to be $P_t = 40dBm$. %Search space dimension $D=12$, population size $N=4$, and the maximum number of iterations $L=10000$.
%\begin{figure}[h]
%\centering
%\includegraphics[width=3.6in]{6.png}
%\caption{Comparison of the sum of energy efficiency of SSA at different IRS shift phase}
%\label{Comparison of the sum of maximum energy efficiency of SSA and SCA at $\theta=\frac{2\pi}{9}$}
%\end{figure}

%From the simulation results in Figure 3, it can be seen that the energy efficiency increases as the number of iterations increases. When the number of iterations is around 40, the energy efficiency can reach a local optimum, especially when the phase shift of IRS $\theta=\frac{2\pi}{9}$, which performs the best.
\begin{figure}[h]
\centering
\includegraphics[width=3in]{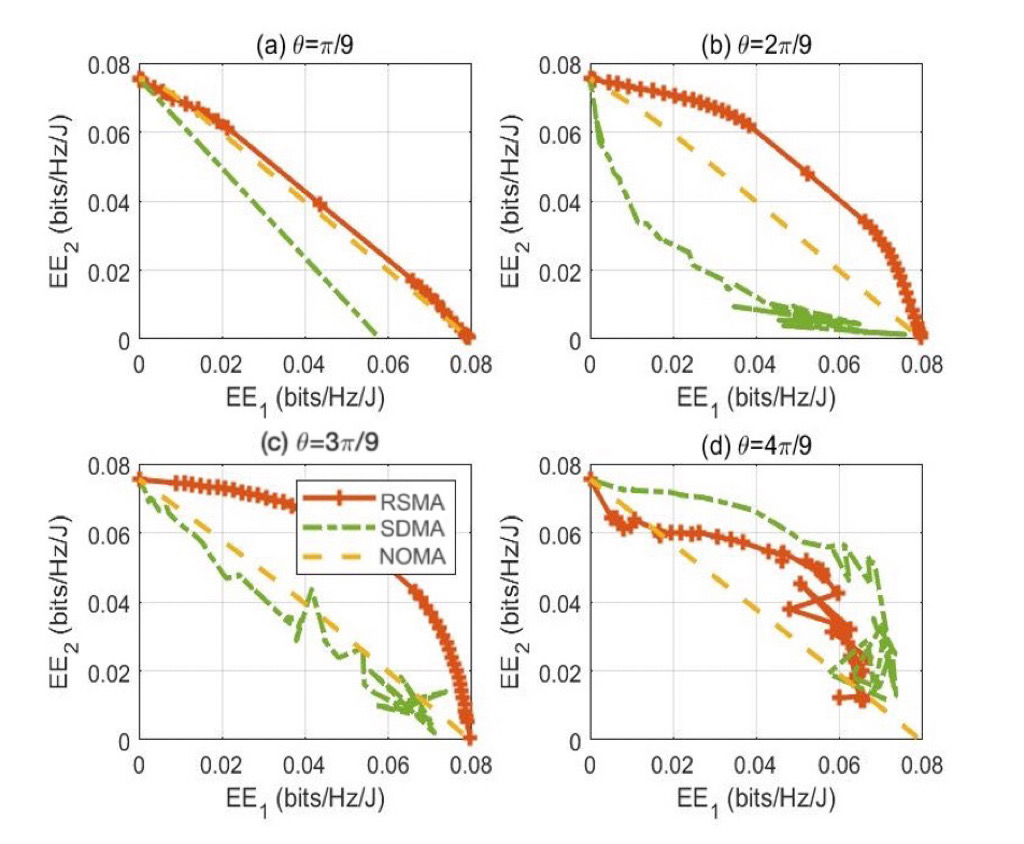}
\caption{Comparison of the energy efficiency of SCA at different IRS phase shift}
%\label{Comparison of the sum of maximum energy efficiency of SSA and SCA at $\theta=\frac{2\pi}{9}$}
\end{figure}
%From the simulation results in Figure 4, it can be seen that when the phase shift of IRS $\theta=\frac{3\pi}{9}$, the performance is the best, followed by $\theta=\frac{2\pi}{9}$. As the number of passive reflective elements of IRS increases, the energy efficiency tends to decrease. %Therefore, this paper chooses to compare the energy efficiency under the SSA and SCA when the phase shift of IRS is $\theta=\frac{2\pi}{9}$.
\begin{figure}[h]
\centering
\includegraphics[width=3in]{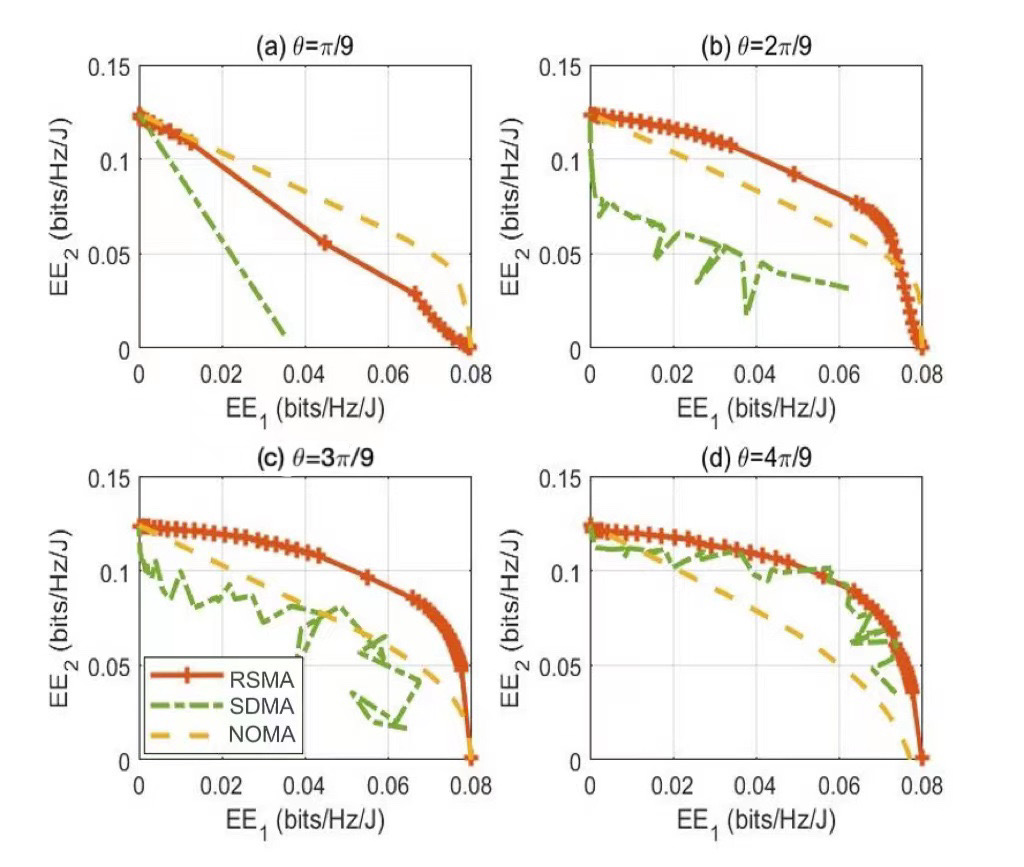}
\caption{Comparison of the energy efficiency of SSA at different IRS phase shift}
%\label{Comparison of the sum of maximum energy efficiency of SSA and SCA at $\theta=\frac{2\pi}{9}$}
\end{figure}
Figures 3 and 4 compare the energy efficiencies of RSMA, NOMA, and SDMA using the SCA and SSA methods at different phase shift angles of the IRS. It can be seen that SSA is much better than SCA in terms of overall energy efficiency optimization, an increase of $60\%$. From each subgraph, we can find that RSMA outperforms NOMA and SDMA, especially when the phase shift angle needs to be orthogonal and aligned with the effective user receiver channel. As $\theta$ increases, the gap between RSMA and SDMA decreases because SDMA works well when the phase shift angle is sufficiently orthogonal to the user channel, where RSMA is significantly better than NOMA. So RSMA achieves more significant energy efficiency than SDMA and NOMA in the case of multiple phase shift angles and efficient user receiver channels.
%Figure 3 and Figure 4 compare the energy efficiency of RSMA, NOMA, and SDMA using SCA and SSA methods at different phase shift angles of IRS, respectively. SSA is much better than SCA in overall energy efficiency optimization, which is improved by $60\%$. From each subplot, we can find that RSMA outperforms NOMA and SDMA in most cases. %Especially when the phase shift of the IRS is $\theta=\frac{2\pi}{9}$, the gap between RSMA and SDMA is the largest.As $\theta$ increases, the gap between RSMA and SDMA decreases. When the phase shift of the IRS is $\theta=\frac{4\pi}{9}$, the SDMA performs well, and the energy efficiency of the RSMA is significantly better than that of the NOMA.

\begin{figure}[h]
\centering
\includegraphics[width=3in]{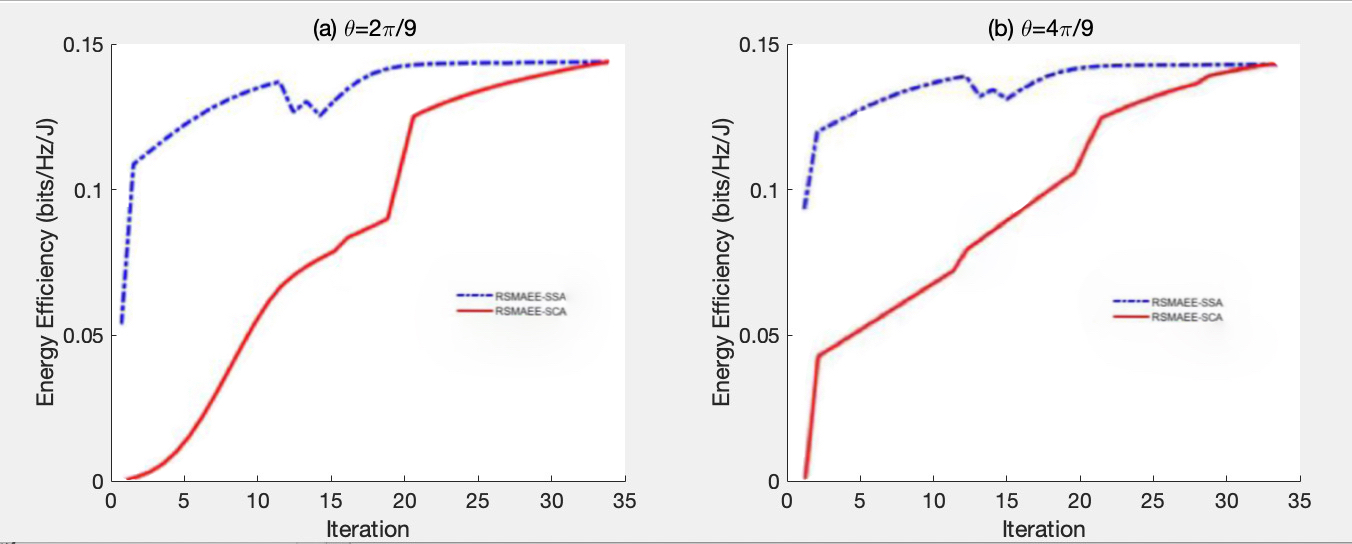}
\caption{Comparison of the maximum energy efficiency of SSA and SCA at $\theta=\frac{2\pi}{9}$ and $\theta=\frac{4\pi}{9}$}
%\label{Comparison of the sum of maximum energy efficiency of SSA and SCA at $\theta=\frac{2\pi}{9}$}
\end{figure}
Figure 5 compares the maximum energy efficiency of the system when the IRS phase shift $\theta=\frac{2\pi}{9}$ and $\theta=\frac{4\pi}{9}$ using the SSA and SCA methods. Compared with SCA, SSA has a significant improvement in energy efficiency. The pictures clearly show that the energy efficiency of both algorithms increases significantly with the increase of the number of iterations under the limit of the total transmit power of the system. In addition, with fewer iterations, SSA can achieve the maximum energy efficiency of the system faster than SCA, regardless of whether the phase shift angle of the IRS is designed to be orthogonal or aligned with the effective user channel.

The progressive time complexity of SSA is $\mathcal{O}(t(d\times n+n))$, where $t$ is the number of iterations, $d$ is the number of variables, $n$ is the scale of the problem. However, the progressive time complexity of of SCA is $\mathcal{O}(td^3)$. Obviously, the progressive time complexity of SCA is much higher than that of SSA. The experimental results show that the energy efficiency of SSA is significantly enhanced compared with SCA, while the time complexity is also reduced considerably. However, the global optimal result to the SSA optimization problem is unknown. In this case, it is assumed that the optimal solution obtained so far is the global optimal solution. 

\section{CONCLUSION}
This paper introduces the rate split multiple access (RSMA) technology into the terahertz band IRS system and focuses on the energy efficiency optimization of the MISO downlink system with two users. From the formula of time complexity, it can be seen that SCA consumes much more computation time than SSA. Simulation results in MATLAB show that RSMA outperforms NOMA and SDMA in terms of energy efficiency. SSA can prevent the results from falling into local optimum points, while SCA can easily approach singular points, resulting in inaccurate calculation results. SSA can effectively improve the system's energy efficiency and consume less time, thereby optimizing the overall performance. At the same time, IRS can intelligently enhance the quality of wireless channels and promote wireless transmission. In our future work, we would like to compare capacities  between the RSMA and NOMA or SDMA systems and analyze the received SINR based on our previous works, such as  \cite{network_capacity,Chen2021MIMO,MIMO_capacity,cellular8,cellular9,cellular10,cellular11}.


\begin{thebibliography}{00}
%\bibitem{1} S.Buzzi,C.L.I,T.E.Klein,H.V.Poor,C.Yang,andA.Zappone,“A survey of energy-efficient techniques for 5G networks and challenges ahead,” IEEE Journal on Selected Areas in Communications, vol. 34, no. 4, pp. 697–709, April 2016.
%\bibitem{2} Tervo,L.N.Tran,andM.Juntti,“Optimalenergy-efficienttransmit beamforming for multi-user MISO downlink,” IEEE Transactions on Signal Processing, vol. 63, no. 20, pp. 5574–5588, Oct 2015.
%\bibitem{3} O. Tervo, A. To ̈lli, M. Juntti, and L. N. Tran, “Energy-efficient beam coordination strategies with rate-dependent processing power,” IEEE Transactions on Signal Processing, vol. 65, no. 22, pp. 6097–6112, Nov 2017. 271--350.
%\bibitem{4} M. Zeng, A. Yadav, O. A. Dobre, G. I. Tsiropoulos, and H. V. Poor, “Capacity comparison between MIMO-NOMA and MIMO-OMA with multiple users in a cluster,” IEEE J. Sel. Areas Commun., vol. 35, no. 10, pp. 2413–2424, Oct. 2017.
%\bibitem{5} Y. Yang, M. Pesavento, S. Chatzinotas, and B. Ottersten, “En- ergy efficiency optimization in MIMO interference channels: A successive pseudoconvex approximation approach,” arXiv preprint arXiv:1802.06750, 2018.
%\bibitem{6} Q. Sun, S. Han, C. L. I, and Z. Pan, “Energy efficiency optimization for fading MIMO non-orthogonal multiple access systems,” in 2015 IEEE International Conference on Communications (ICC), June 2015, pp. 2668–2673.
%\bibitem{7} P. Wu, J. Zeng, X. Su, H. Gao, and T. Lv, “On energy efficiency optimization in downlink MIMO–NOMA,” in 2017 IEEE International Conference on Communications Workshops (ICC Workshops), May 2017, pp. 399–404.
%\bibitem{8} Y. Mao, B. Clerckx, and V. O. Li, “Rate-splitting multiple access for downlink communication systems: bridging, generalizing, and outperforming SDMA and NOMA,” EURASIP Journal on Wireless Communications and Networking, vol. 2018, no. 1, p. 133, May 2018.
%\bibitem{9} A. Zappone, B. Matthiesen, and E. A. Jorswieck, “Energy efficiency in MIMO underlay and overlay device-to-device communications and cognitive radio systems,” IEEE Transactions on Signal Processing, vol. 65, no. 4, pp. 1026–1041, Feb 2017.
%\bibitem{12} S. He, Y. Huang, J. Wang, L. Yang, and W. Hong, “Joint antenna selection and energy-efficient beamforming design,” IEEE Signal Pro- cessing Letters, vol. 23, no. 9, pp. 1165–1169, Sept 2016.
%\bibitem{13} H.Q.Ngo,L.N.Tran,T.Q.Duong,M.Matthaiou,andE.G.Larsson, “On the total energy efficiency of cell-free massive MIMO,” IEEE Transactions on Green Communications and Networking, vol. 2, no. 1, pp. 25–39, March 2018.
\bibitem{du2022ITS_RIS} W. Du, G. Chen, P. Xiao, Z. Lin, C. Huang. W. Hao, and R. Tafazolli, "Weighted Sum-Rate and Energy Efficiency Maximization for Joint ITS and IRS Assisted Multiuser MIMO Networks," in IEEE Transactions on Communications, 2022, doi: 10.1109/TCOMM.2022.3213356.

\bibitem{chu2022IoT_RIS} Z. Chu, P. Xiao, D. Mi, W. Hao, Z. Lin, Q. Chen, and R. Tafazolli, “Wireless powered intelligent radio environment with non-linear
energy harvesting,” IEEE Internet of Things Journal, pp. 1–1, 2022.

\bibitem{Hu2021CodedRIS} Y. Hu, P. Wang, Z. Lin, and M. Ding, “Performance analysis of
reconfigurable intelligent surface assisted wireless system with low-
density parity-check code,” IEEE Communications Letters, vol. 25,
no. 9, pp. 2879–2883, 2021.


\bibitem{xu2021THzUAV} X. Wang, P. Wang, M. Ding, Z. Lin, F. Lin, B. Vucetic, and L. Hanzo, “Performance analysis of terahertz unmanned aerial vehicular net-
works,” IEEE Transactions on Vehicular Technology, vol. 69, no. 12,
pp. 16 330–16 335, 2020.
\bibitem{xu2022THzRIS} X. Wang, Z. Lin, F. Lin, and L. Hanzo, “Joint hybrid 3d beamforming relying on sensor-based training for reconfigurable intelligent surface
aided terahertz-based multi-user massive mimo systems,” IEEE Sen-
sors Journal, pp. 1–1, 2022.
\bibitem{Liu2022THzRIS} Y. Liu, W. Li, Z. Lin, "A Dynamic Subarray Structure in Reconfigurable Intelligent Surfaces for TeraHertz Communication Systems", Proceedings of IEEE Conference on Standards for Communications and Networking (CSCN), November 2022, Thessaloniki, Greece.
\bibitem{1}M. Zeng, A. Yadav, O. A. Dobre, G. I. Tsiropoulos, and H. V. Poor, “Capacity comparison between MIMO-NOMA and MIMO-OMA with multiple users in a cluster,” IEEE J. Sel. Areas Commun., vol. 35, no. 10, pp. 2413–2424, Oct 2017.
\bibitem{2}Y. Mao, B. Clerckx, and V. O. Li, “Rate-splitting multiple access for downlink communication systems: bridging, generalizing, and outperforming SDMA and NOMA,” EURASIP Journal on Wireless Communications and Networking, vol. 2018, no. 1, p. 133, May 2018.
\bibitem{3}B. Clerckx, Y. Mao, R. Schober and H. V. Poor, "Rate-Splitting Unifying SDMA, OMA, NOMA, and Multicasting in MISO Broadcast Channel: A Simple Two-User Rate Analysis," in IEEE Wireless Communications Letters, vol. 9, no. 3, pp. 349-353, March 2020, doi: 10.1109/LWC.2019.2954518.
\bibitem{4}L. Yin, O. Dizdar and B. Clerckx, "Rate-Splitting Multiple Access for Multigroup Multicast Cellular and Satellite Communications: PHY Layer Design and Link-Level Simulations," 2021 IEEE International Conference on Communications Workshops (ICC Workshops), 2021, pp. 1-6, June 2021.
\bibitem{5}O. Dizdar, Y. Mao, Y. Xu, P. Zhu and B. Clerckx, "Rate-Splitting Multiple Access for Enhanced URLLC and eMBB in 6G: Invited Paper," 2021 17th International Symposium on Wireless Communication Systems (ISWCS), 2021, pp. 1-6,Sept 2021.
\bibitem{6}A. Mishra, Y. Mao, O. Dizdar and B. Clerckx, "Rate-Splitting Multiple Access for Downlink Multiuser MIMO: Precoder Optimization and PHY-Layer Design," in IEEE Transactions on Communications, vol. 70, no. 2, pp. 874-890, Feb 2022.
\bibitem{7}H. Fu, S. Feng and D. W. Kwan Ng, "Resource Allocation Design for IRS-Aided Downlink MU-MISO RSMA Systems," 2021 IEEE International Conference on Communications Workshops (ICC Workshops), 2021, pp. 1-6, June 2021.
\bibitem{8}Y. Mao, B. Clerckx and V. O. K. Li, "Energy Efficiency of Rate-Splitting Multiple Access, and Performance Benefits over SDMA and NOMA," 2018 15th International Symposium on Wireless Communication Systems (ISWCS), 2018, pp. 1-5, Aug 2018.
\bibitem{9}Y. Yang, M. Pesavento, S. Chatzinotas, and B. Ottersten, “Energy efficiency optimization in MIMO interference channels: A successive pseudoconvex approximation approach,” arXiv preprint arXiv:1802.06750, 2018.
\bibitem{10}A. -A. A. Boulogeorgos, E. N. Papasotiriou and A. Alexiou, "A Distance and Bandwidth Dependent Adaptive Modulation Scheme for THz Communications," 2018 IEEE 19th International Workshop on Signal Processing Advances in Wireless Communications (SPAWC), 2018, pp. 1-5, doi: 10.1109/SPAWC.2018.8445864.
%\bibitem{12} Q. Sun, S. Han, C. L. I, and Z. Pan, “Energy efficiency optimization for fading MIMO non-orthogonal multiple access systems,” in 2015 IEEE International Conference on Communications (ICC), June 2015, pp. 2668–2673.
%\bibitem{13} P. Wu, J. Zeng, X. Su, H. Gao, and T. Lv, “On energy efficiency optimization in downlink MIMO–NOMA,” in 2017 IEEE International Conference on Communications Workshops (ICC Workshops), May 2017, pp. 399–404.
\bibitem{11}S. Buzzi, C. -L. I, T. E. Klein, H. V. Poor, C. Yang and A. Zappone, "A Survey of Energy-Efficient Techniques for 5G Networks and Challenges Ahead," in IEEE Journal on Selected Areas in Communications, vol. 34, no. 4, pp. 697-709, April 2016, doi: 10.1109/JSAC.2016.2550338.
\bibitem{12}Y. Pan, K. Wang, C. Pan, H. Zhu, and J. Wang, “SUM-rate maximization for intelligent reflecting surface assisted Terahertz Communications,” IEEE Transactions on Vehicular Technology, vol. 71, no. 3, pp. 3320–3325, 2022.
\bibitem{13}W. Tang et al., "Path Loss Modeling and Measurements for Reconfigurable Intelligent Surfaces in the Millimeter-Wave Frequency Band," in IEEE Transactions on Communications, vol. 70, no. 9, pp. 6259-6276, Sept. 2022, doi: 10.1109/TCOMM.2022.3193400.
\bibitem{14}O. Tervo, L. -N. Tran and M. Juntti, "Optimal Energy-Efficient Transmit Beamforming for Multi-User MISO Downlink," in IEEE Transactions on Signal Processing, vol. 63, no. 20, pp. 5574-5588, Oct.15, 2015, doi: 10.1109/TSP.2015.2453134.
\bibitem{15}O. Tervo, et. al, "Energy-efficient beam coordination strategies with rate dependent processing power," IEEE Transactions on Signal Processing, vol. 65, no. 22, pp. 6097–6112, Nov 2017. 271--350.
\bibitem{16}S. He, Y. Huang, J. Wang, L. Yang, and W. Hong, “Joint antenna selection and energy-efficient beamforming design,” IEEE Signal Pro- cessing Letters, vol. 23, no. 9, pp. 1165–1169, Sept 2016.
\bibitem{17}H. Q. Ngo, L. -N. Tran, T. Q. Duong, M. Matthaiou and E. G. Larsson, "On the Total Energy Efficiency of Cell-Free Massive MIMO," in IEEE Transactions on Green Communications and Networking, vol. 2, no. 1, pp. 25-39, March 2018, doi: 10.1109/TGCN.2017.2770215.

\bibitem{network_capacity}G. Mao, Z. Lin, X. Ge, Y. Yang, "Towards a simple relationship to estimate the capacity of static and mobile wireless networks", IEEE transactions on wireless communications 12 (8), 2014, 3883-3895	
\bibitem{Chen2021MIMO} Y. Chen, M. Ding, D. L opez-Perez, X. Yao, Z. Lin, and G. Mao, "On the theoretical analysis of network-wide massive mimo performance
and pilot contamination," IEEE Transactions on Wireless Communi-
cations, vol. 21, no. 2, pp. 1077–1091, 2022.
\bibitem{MIMO_capacity}Z. Lin, B. Vucetic, J. Mao, "Ergodic capacity of LTE downlink multiuser MIMO systems", 2008 IEEE International Conference on Communications, 3345-3349.

\bibitem{cellular8}Z. Lin, P. Xiao and B. Vucetic, "SINR distribution for LTE downlink multiuser MIMO systems," 2009 IEEE International Conference on Acoustics, Speech and Signal Processing, 2009, pp. 2833-2836, doi: 10.1109/ICASSP.2009.4960213.

\bibitem{cellular9}Z. Lin, T. B. Sorensen and P. E. Mogensen, "Downlink SINR Distribution of Linearly Precoded Multiuser MIMO Systems," in IEEE Communications Letters, vol. 11, no. 11, pp. 850-852, November 2007, doi: 10.1109/LCOMM.2007.071082.

\bibitem{cellular10}M. Ding, D. López-Pérez, G. Mao and Z. Lin, "Microscopic Analysis of the Uplink Interference in FDMA Small Cell Networks," in IEEE Transactions on Wireless Communications, vol. 15, no. 6, pp. 4277-4291, June 2016, doi: 10.1109/TWC.2016.2538261.

\bibitem{cellular11}J. Yang, M. Ding, G. Mao and Z. Lin, "Interference Management in In-Band D2D Underlaid Cellular Networks," in IEEE Transactions on Cognitive Communications and Networking, vol. 5, no. 4, pp. 873-885, Dec. 2019, doi: 10.1109/TCCN.2019.2927568.

\end{thebibliography}
\end{document}